\documentclass[preprint,3p,12pt]{elsarticle}
\usepackage{setspace}
\usepackage{lineno,hyperref}
\usepackage{amsmath, amssymb, amsfonts, bm}
\usepackage{float}
\usepackage{mathtools}
\usepackage{color}
\usepackage{soul}
\usepackage{xcolor}
\usepackage{physics}

\usepackage{caption}
\journal{Elsevier}

\begin{document}
	\begin{frontmatter}
		
		\title{Stiff deployable structures via coupling of thick Miura-ori tubes along creases}

		\author[inst1]{Sunao Tomita}
		\affiliation[inst1]{organization={Toyota Central R\&D Labs., Inc.},
			city={Bunkyo-ku},
			state={Tokyo},
			postcode={112-0004}, 
			country={Japan}}
		
		\author[inst1]{Kento Shimanuki}
		\author[inst1]{Kazuhiko Umemoto}
		\author[inst1]{Atsushi Kawamoto}
		\author[inst1]{Tsuyoshi Nomura}

		\author[inst2]{Tomohiro Tachi\corref{mycorrespondingauthor}}
		\cortext[mycorrespondingauthor]{Corresponding author}
		\ead{tachi@idea.c.u-tokyo.ac.jp}
		\affiliation[inst2]{organization={Department of General Systems Studies, Graduate School of Arts and Sciences, The University of Tokyo},
			addressline={Meguro-ku}, 
			city ={Meguro-ku},
			postcode={153-8902}, 
			state={Tokyo},
			country={Japan}}
		
		\begin{abstract}
			Origami-based structures play an important role in the realization of deployable mechanisms and unique mechanical properties via programmable deformation by folding. 
			Among origami-based structures, tessellation by the coupling of origami tubes enriches the variations in geometry and mechanical properties. 
			However, thickness accommodation is a critical problem in engineering applications involving the coupling of thick origami tubes. 
			To solve this problem, this study proposes the coupling of thick Miura-ori tubes along the creases for facile fabrication, which sustains the one-degree-of-freedom (DOF) motion of thick Miura-ori tubes owing to the local mirror symmetry around the coupling interfaces.
			Furthermore, the coupling method contributes to the high stiffness of the coupled Miura-ori tubes, as evidenced by the wide gap in the eigenvalues between the one-DOF mode and the elastic modes obtained by the bar-and-hinge models.
			Finally, meter-scale coupled Miura-ori tubes were fabricated to demonstrate one-DOF motion and high stiffness.
			The findings of this study enable the rapid construction of structures by one-DOF motion and enhancement of transportability via flat-foldability.
		\end{abstract}
		
	\end{frontmatter}
	\newpage
	\doublespacing

	\section{Introduction}
	Rigid origami, in which rigid panels are connected via creases, has a significant advantage in realizing folding and deployment motions. Because programmable motions by folding are scalable, rigid origami has potential applications at various scales, from the deployment of solar sails in space \cite{RN390} to the construction of architectural structures \cite{RN475}, shelters \cite{RN56}, arches \cite{RN474}, and medical stents \cite{RN244}. Furthermore, it has been used to produce sandwich panel cores by bending aluminum sheets \cite{RN484}. More recently, self-folding technology has been developed to automatically fold origami-based structures using external stimulation \cite{RN53,RN480}, and has been applied to the deformation of robotics \cite{RN482}.
	
	In addition to the generation of folding and deployment motions, mechanical metamaterials using coupling origami tubes have been considered to obtain unique mechanical properties. In previous studies, origami metamaterials with negative Poisson's ratios \cite{RN365}, multistable cellular structures \cite{RN366}, programmable collapse \cite{RN485}, and extremely stiff deployable structures, referred to as zipper-coupled tubes \cite{RN13,RN253}, have been proposed based on the Miura ori. Tachi-Miura polyhedra are also composed of origami tubes \cite{RN81,RN80,RN64}, which lead to auxeticity \cite{RN60}, load-bearing capability \cite{RN19}, and high-efficiency energy absorption \cite{RN258,RN464}. In these origami structures, thickness accommodation has not been treated to ensure the mechanisms of thick origami structures for flat foldability.
	
	In contrast, engineering applications of rigid origami require thickness accommodation because the ideal origami theory does not apply to the mechanisms of origami with finite thickness.
	In previous studies, the trimming of thick panels \cite{RN264} and the shifting of hinge locations \cite{RN226,RN194} were proposed to avoid interference during folding. In addition, accommodations with various thicknesses for thick origami were proposed and reviewed by Lang et al.\cite{RN225}. Thick origami designs are used for various patterns, such as Miura-ori\cite{RN483,RN223}, Resch patterns \cite{RN177}, and waterbombs \cite{RN398}. In addition, deployments of spheres \cite{RN222}, tubular structures \cite{RN191,RN249}, and dome shapes \cite{RN457} were designed using thick origami. Furthermore, a method for simulating the folding of thick origami was proposed \cite{RN469,RN459}. However, origami-tessellated structures with thick panels have not been fully established.
	
	This study proposes coupled thick Miura-ori tubes along the creases to retain the one-DOF deformation by local mirror symmetry around the coupling interfaces. To this end, Section \ref{sec:Geometry} organizes the coupling of Miura-ori tubes to clarify deployable and undeployable coupled thick Miura-ori tubes. 
	Consequently, the coupling of mirrored Miura-ori tubes was observed to be a new coupling of Miura-ori tubes that maintains the deployable motion of coupled thick Miura-ori tubes.
	In addition, rotated Miura-ori tubes provide thickness accommodation for tessellation structures consisting of Miura-ori tubes, known as interleaved origami cellular structures \cite{RN11}. 
	Furthermore, Section~\ref{sec:Elastic deformation} presents the eigenvalue analysis of bar-and-hinge models \cite{RN196,RN17}, which resulted in mirrored and rotated Miura-ori tubes that increase the stiffness of the coupled Miura-ori tubes by two orders of magnitude after coupling, similar to the zipper coupled tubes known as the stiffest origami tubular structures \cite{RN13}.
	For demonstration, the construction of a meter-scale pillar using coupled thick mirrored Miura-ori tubes based on the one-DOF motion and the high stiffness of the coupled Miura-ori tubes is presented in Section~\ref{sec:fabrication}. Finally, conclusions are presented in Section~\ref{Sec:Conclusion}.
	
	
	\section{Geometry}
	\label{sec:Geometry}
	
	\subsection{Miura-ori tube}
\begin{figure}[H]
		\centering\includegraphics[width=1\linewidth]{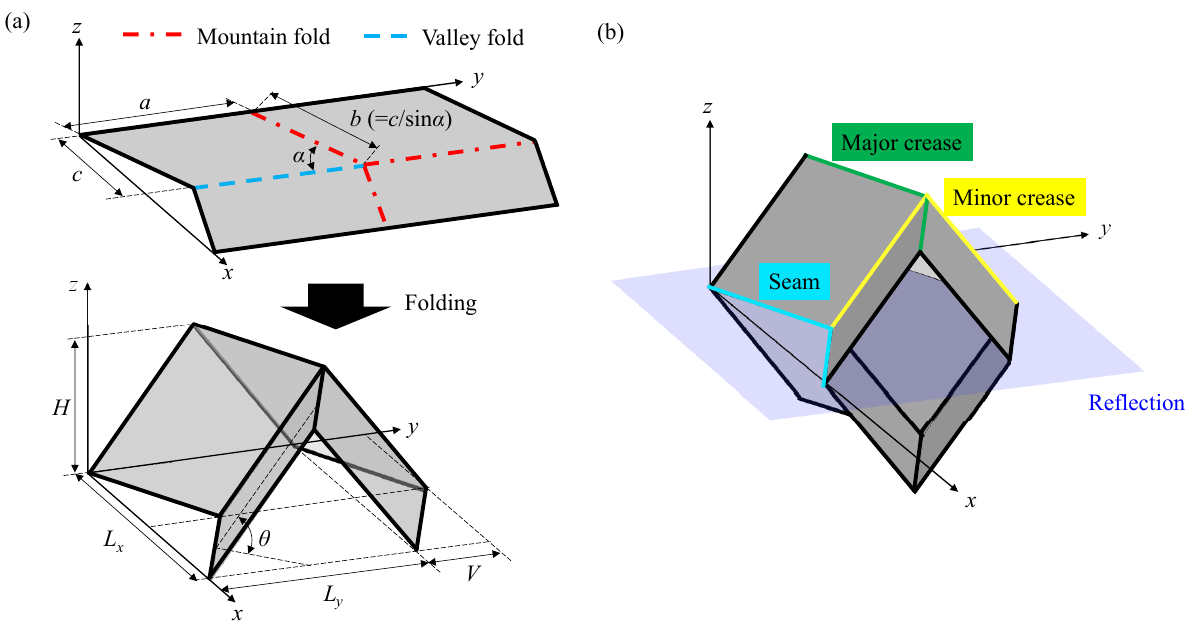}
		\caption{Geometry of Miura-ori tube (a) Unit cell of Miura-ori sheet (b) Miura-ori tube created by coupling mirrored Miura-ori sheet.}
		\label{fig:Miura tube}
	\end{figure}

	The dimensions of a Miura-ori unit cells are indicated in Fig.~\ref{fig:Miura tube}(a). Using the length of Miura-ori $a$, $b$, the internal angle of the parallelogram $\alpha$, and the half dihedral angle $\theta \in [0,\pi/2]$, the dimensions of the unit cells are defined as follows:
	
	\begin{equation}
		L_{x}=2b\frac{\cos\theta\tan\alpha}{\sqrt{1+\cos^2\theta\tan^2\alpha}},
		\label{eq:Lx}
	\end{equation}
	
	\begin{equation}
		L_{y}=2a\sqrt{1-\sin^2\theta\sin^2\alpha},
		\label{eq:Ly}
	\end{equation}
	
	\begin{equation}
		H=2a\sin\theta\sin\alpha,
		\label{eq:H}
	\end{equation}
	and
	\begin{equation}
		V=b\frac{1}{\sqrt{1+\cos^2\theta\tan^2\alpha}}.
		\label{eq:V}
	\end{equation}
	
	Miura-ori tubes were created by coupling Miura-ori to those mirrored along the $x$–-$y$ plane, as shown in Fig.~\ref{fig:Miura tube}(b). Miura-ori tubes have three types of folding lines: i) major crease, ii) minor crease, and iii) seam.
	
	\subsection{Origami tessellation by Miura-ori tubes}
	\label{sec:tessellation by thin Miura-ori}
\begin{figure}[H]
		\centering\includegraphics[width=0.9\linewidth]{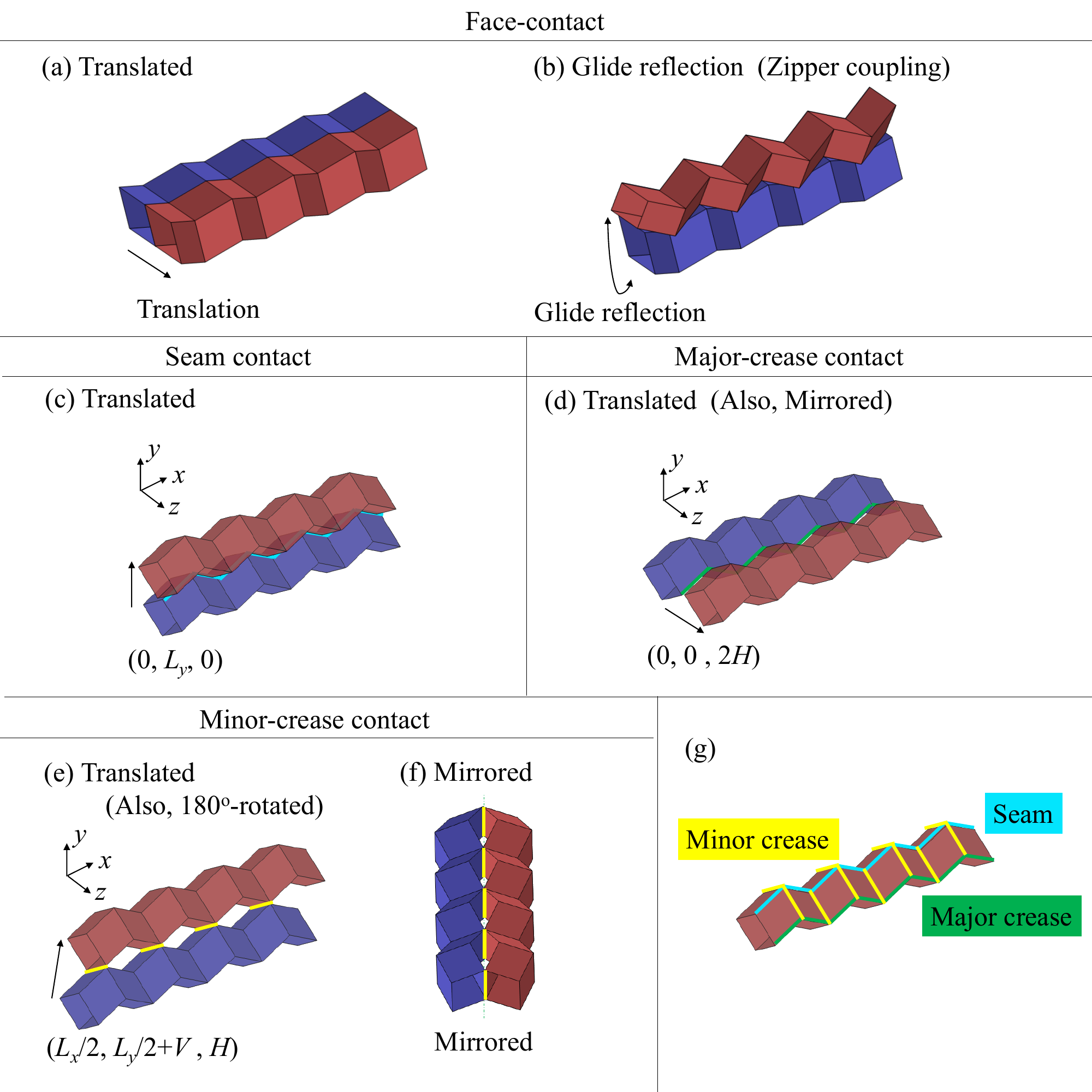}
		\caption{Classification of the coupling of Miura-ori tubes. (a) Face contact by translation. (b) Face contact by glide reflection (i.e., zipper-coupled tubes). (c) Seam contact by translation. (d) Major-creases contact by translation. (e) Minor-creases contact by translation. (f) Minor-creases contact by mirror. (g) Classification of the edges of Miura-ori tubes.}
		\label{fig:Miura Coupling}
	\end{figure}

\begin{figure}[H]
		\centering\includegraphics[width=0.9\linewidth]{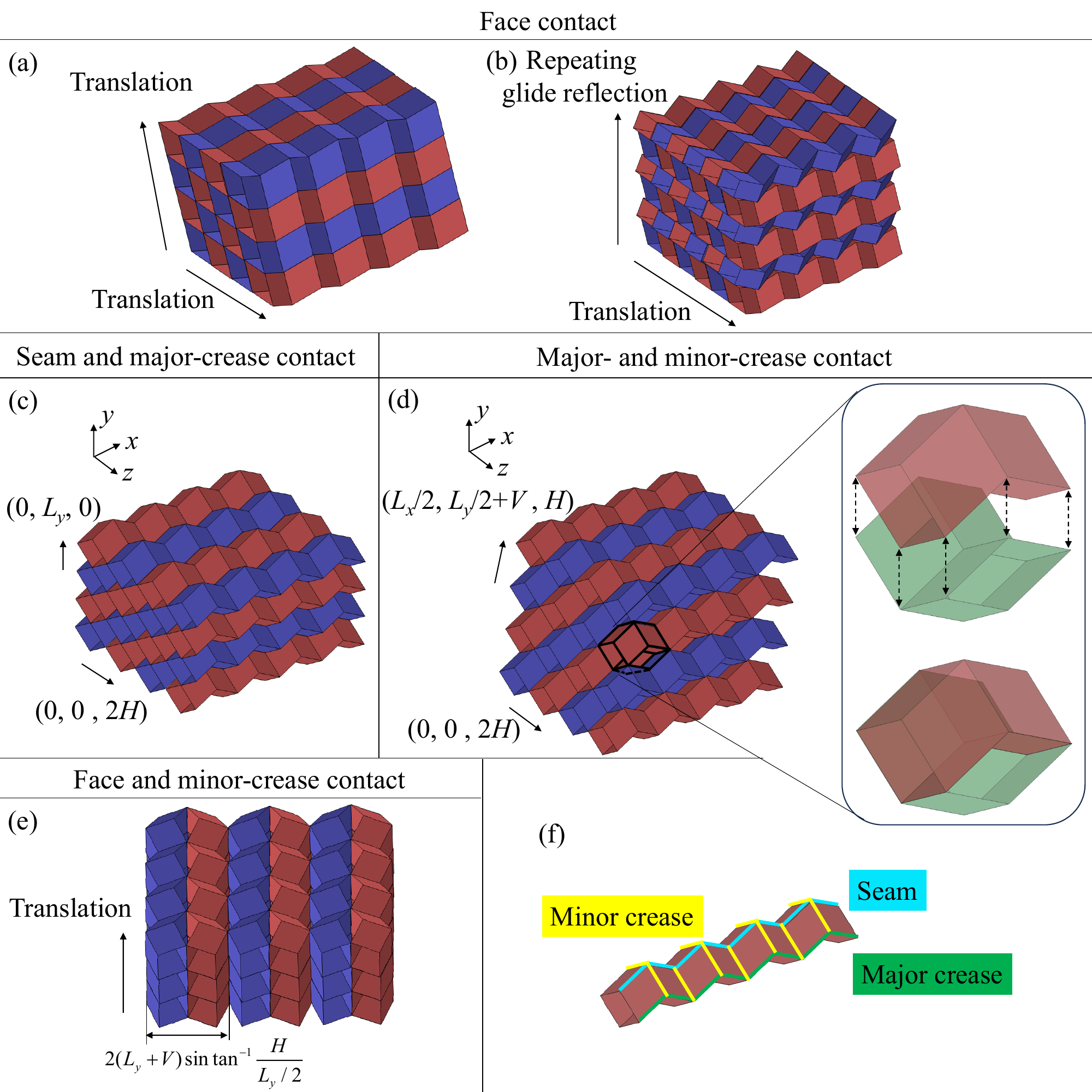}
		\caption{Tessellation by coupling of the Miura-ori tubes. (a) Tessellation of translated Miura-ori tubes by face contact. (b) Tessellation of zipper-coupled Miura-ori tubes by face contact. (c) Combination of seam and major-crease contact. (d) Combination of major- and major-crease contacts. (e) Combination of face and minor-crease contacts. (f) Classification of edges of Miura-ori tubes.}
		\label{fig:Miura tessellation}
	\end{figure}

	This subsection describes tessellation via coupling of the Miura-ori tubes by first classifying the coupling into couplings along the faces and creases in Fig.~\ref{fig:Miura Coupling} and discuss the spatial tessellation achieved by their combination in Fig.~\ref{fig:Miura tessellation}.
	
	Translated and rotated Miura-ori tubes can be coupled along the faces, as depicted in Fig.~\ref{fig:Miura Coupling}(a) and (b).
	The coupling of the rotated Miura-ori tubes along the faces leads to glide reflection, which is referred to as a zipper-coupled tube \cite{RN13}. 
	In addition to coupling along the faces, Miura-ori tubes can also be coupled along their creases and seams. 
	Seam contact based on translation along the vector $(0,{L_y},0)$ yields the coupled structures shown in Fig.~\ref{fig:Miura Coupling}(c). 
	In addition, the major-crease contact leads to coupled Miura-ori tubes, as shown in Fig.~\ref{fig:Miura Coupling}(d) based on the translation vector $(0,0,2H)$, which is also generated by a mirror reflection along the major crease. 
	The minor-crease contact also provides two types of coupled Miura-ori tubes, as shown in Fig.~\ref{fig:Miura Coupling}(e) and (f), respectively. 
	The coupled Miura-ori tubes in Fig.~\ref{fig:Miura Coupling}(e) exhibit $180^\circ$rotations or translations along the vector $({L_x}/2,{L_y}/2+V,H)$.
	Furthermore, the Miura-ori tubes mirrored by the minor-crease contact can be coupled, as indicated in Fig.~\ref{fig:Miura Coupling}(f). Classification of the edges of Miura-ori tubes is depicted in Fig.~\ref{fig:Miura Coupling}(g).
	
	With these coupled Miura-ori tubes, the tessellation of Miura-ori tubes produces cellular structures.
	Three-dimensional (3D) tessellation was obtained by translating the face-contact translated tubes along one of the minor-crease directions (Fig.~\ref{fig:Miura tessellation}(a)).
	Another round of 3D tessellation was obtained by translating the zipper coupling along the minor-crease direction (Fig.~\ref{fig:Miura tessellation}(b)).
	
	The combination of seam and major-crease contacts (i.e., translations along the vectors $(0,{L_y},0)$ and $(0,0,2H)$) leads to the cellular structures shown in Fig.~\ref{fig:Miura tessellation}(c). 
	This tessellation is equivalent to coupling along the faces, as shown in Fig.~\ref{fig:Miura tessellation}(a), because the gaps between Miura-ori tubes form other Miura-ori tubes. In addition, the combination of major- and minor-crease contacts (i.e., translations along the vectors $(0,0,2H)$ and $({L_x}/2,{L_y}/2+V,H)$) leads to the cellular structures shown in Fig.~\ref{fig:Miura tessellation}(d), where the gaps between Miura-ori tubes form more Miura-ori tubes in the perpendicular direction, and the cellular assembly is equivalent to interleaved Miura-ori tubes \cite{RN11}.
	Moreover, mirrored Miura-ori tubes can be periodically coupled by a combination of mirror coupling and face contact by repeating the mirror coupling of minor creases to obtain two-dimensional (2D) tessellation and then translating along the minor crease axes to obtain 3D tessellation, as shown in Fig.~\ref{fig:Miura tessellation}(e).
	Mirror coupling to form a 2D tessellation is equivalent to a translation of length $2({L_y}+V)\sin\tan^{-1}\frac{H}{{L_y}/2}$ along the normal vector of the plane of mirror symmetry. Classification of the edges in Miura-ori tubes is shown in Fig.~\ref{fig:Miura tessellation}(f).
	
	\subsection{Thick Miura-ori tubes}
	\label{sec:Miura-ori}
\begin{figure}[H]
		\centering\includegraphics[width=0.8\linewidth]{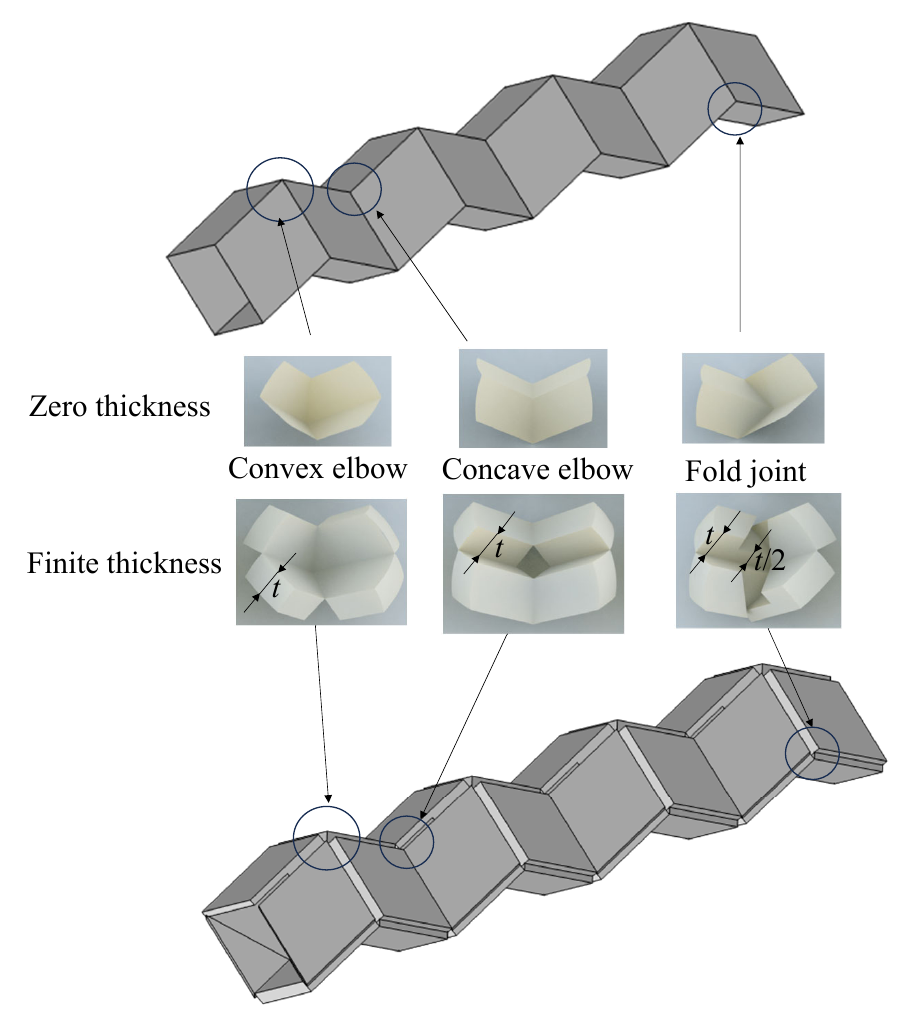}
		\caption{Geometry of a thick Miura-ori tube. The axis of rotation on the folding lines are shifted to avoid inference during folding.}
		\label{fig:Thick Miura tube}
	\end{figure}

	To avoid interference during the folding of the thick Miura-ori, each axis of rotation had to be shifted \cite{RN226}. 
	Mechanisms in thick Miura-ori tubes can be classified into three types (1.Convex elbow, 2.Concave elbow and 3.Fold joint) by thickness accommodation, as shown in Fig.~\ref{fig:Thick Miura tube}. 
	A concave elbow (i.e., an elbow with a negative Gaussian area) and convex elbow (i.e., an elbow with a positive Gaussian area) can be replaced by panels of constant thickness $t$.
	In contrast, a fold joint is replaced by panels with two thicknesses to achieve a valid one-DOF motion, and the thickness of the connection part in the fold joint is set to half of the thickness in the other parts~\cite{RN191}.
	
	\subsection{Coupled Miura-ori tubes along their creases}
	\label{sec:Coupled}
\begin{figure}[H]
		\centering\includegraphics[width=1\linewidth]{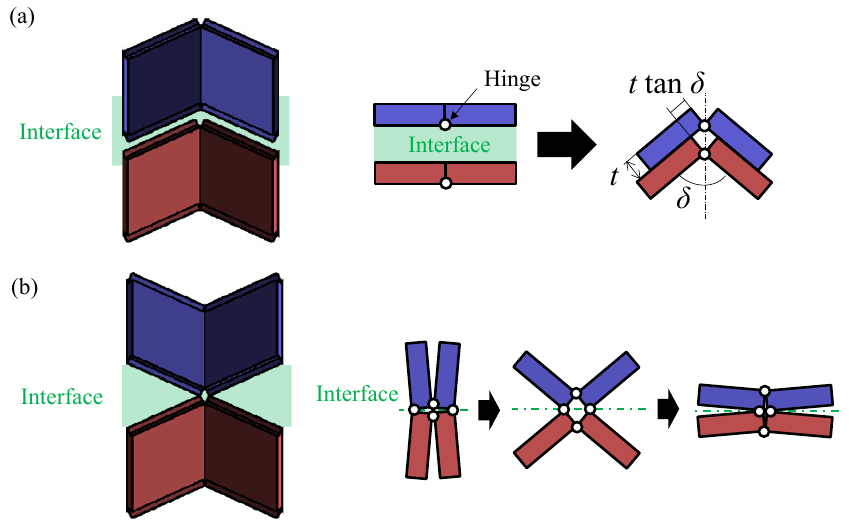}
		\caption{Interface for the coupling of the Miura-ori tubes (a) Coupling along the faces. 
			The thickness of the panels constrains the mechanisms. (b) Coupling along the creases. Local mirror symmetry around interfaces keeps mechanisms after coupling.}
		\label{fig:Motion constraint}
	\end{figure}

	Coupling of thick Miura-ori tubes along the faces is impossible because coupling along the faces produces a relative motion of $t\tan\delta$ because of the misalignment of the folding axis caused by the plate thickness $t$ during folding defined by the angle $\delta$, which blocks the one-DOF motion after coupling, as shown in Fig.~\ref{fig:Motion constraint}(a).
	In contrast, coupling along the edges is possible when the coupled panels are local mirror reflections with respect to a plane. 
	This mechanism was maintained after coupling, as shown in Fig.~\ref{fig:Motion constraint}(b), with scissor-like motions at the coupling interfaces. 
	
\begin{figure}[H]
		\centering\includegraphics[width=1\linewidth]{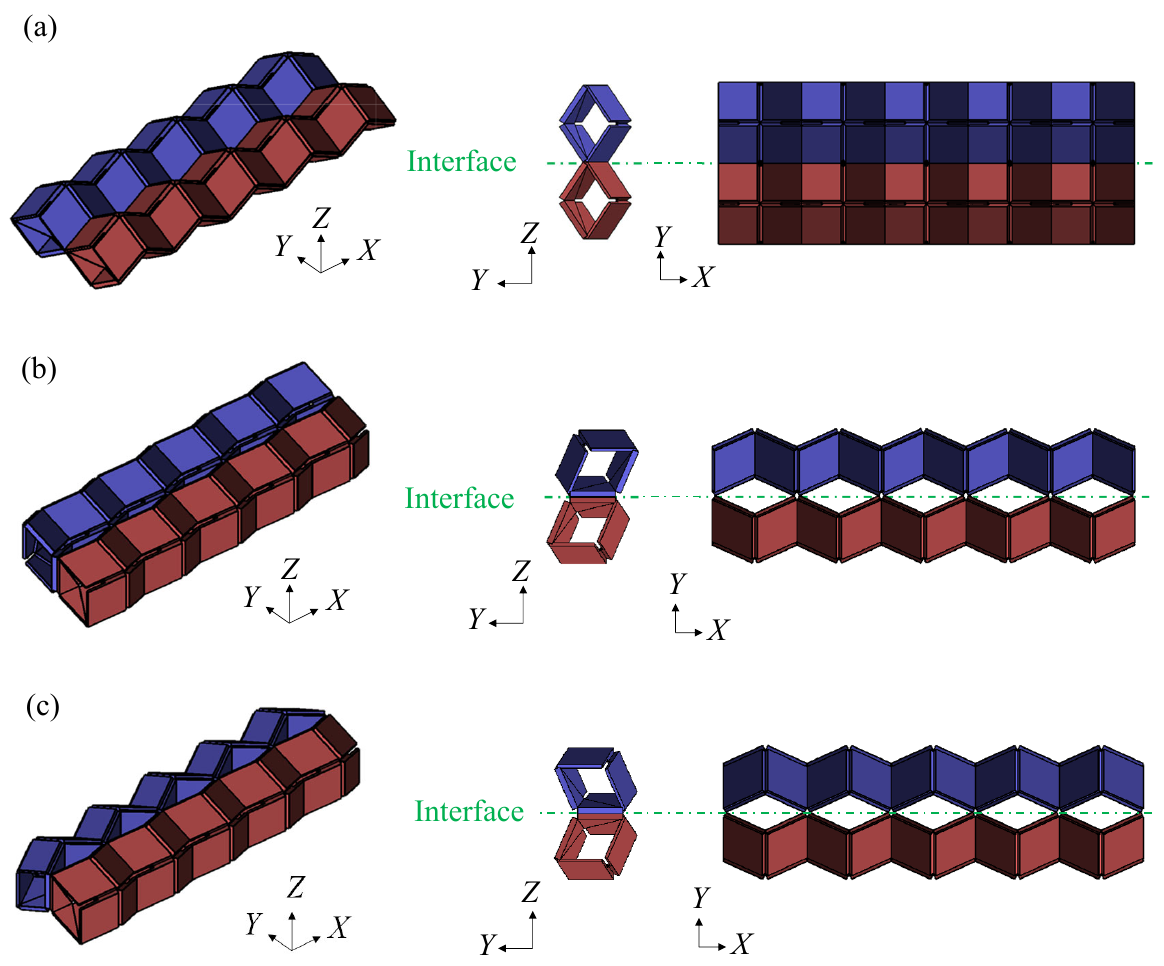}
		\caption{Coupling of thick Miura-ori tubes along creases with one-DOF motions (a) Coupling of translated Miura-ori tubes by major-crease contact. (b) Coupling of rotated Miura-ori tubes by minor-crease contact. (c) Coupling of mirrored Miura-ori tubes by minor-crease contact.}
		\label{fig:Thick Crease coupling}
	\end{figure}

	As indicated in Fig.~\ref{fig:Thick Crease coupling}(a), (b), and (c), Miura-ori tubes can be coupled along the major and minor creases ($180^\circ$ rotation and mirrored), which correspond to the coupling of thin Miura-ori tubes shown in Fig.~\ref{fig:Miura Coupling}(d), (e) and (f). As these coupled Miura-ori tubes ensure local mirror symmetry around the coupled interfaces, if the thicknesses of the panels on the interfaces are the same, the mechanisms can be maintained after coupling the Miura-ori tubes.
	The detailed motion of the coupled thick Miura-ori tubes is shown in Fig.~\ref{fig:Motion of Translated tubes}, \ref{fig:Motion of rotated tubes} and \ref{fig:Motion of mirrored tubes}.
	Note that the coupling of thick Miura-ori tubes along the seam is impossible for reasons similar to those of face contact.
	

	\subsection{Tessellation by thick Miura-ori tubes}
	\label{sec:Tessellation}
	
	\subsubsection{Tessellation of translated tubes in two-dimensional space}
\begin{figure}[H]
		\centering\includegraphics[width=1\linewidth]{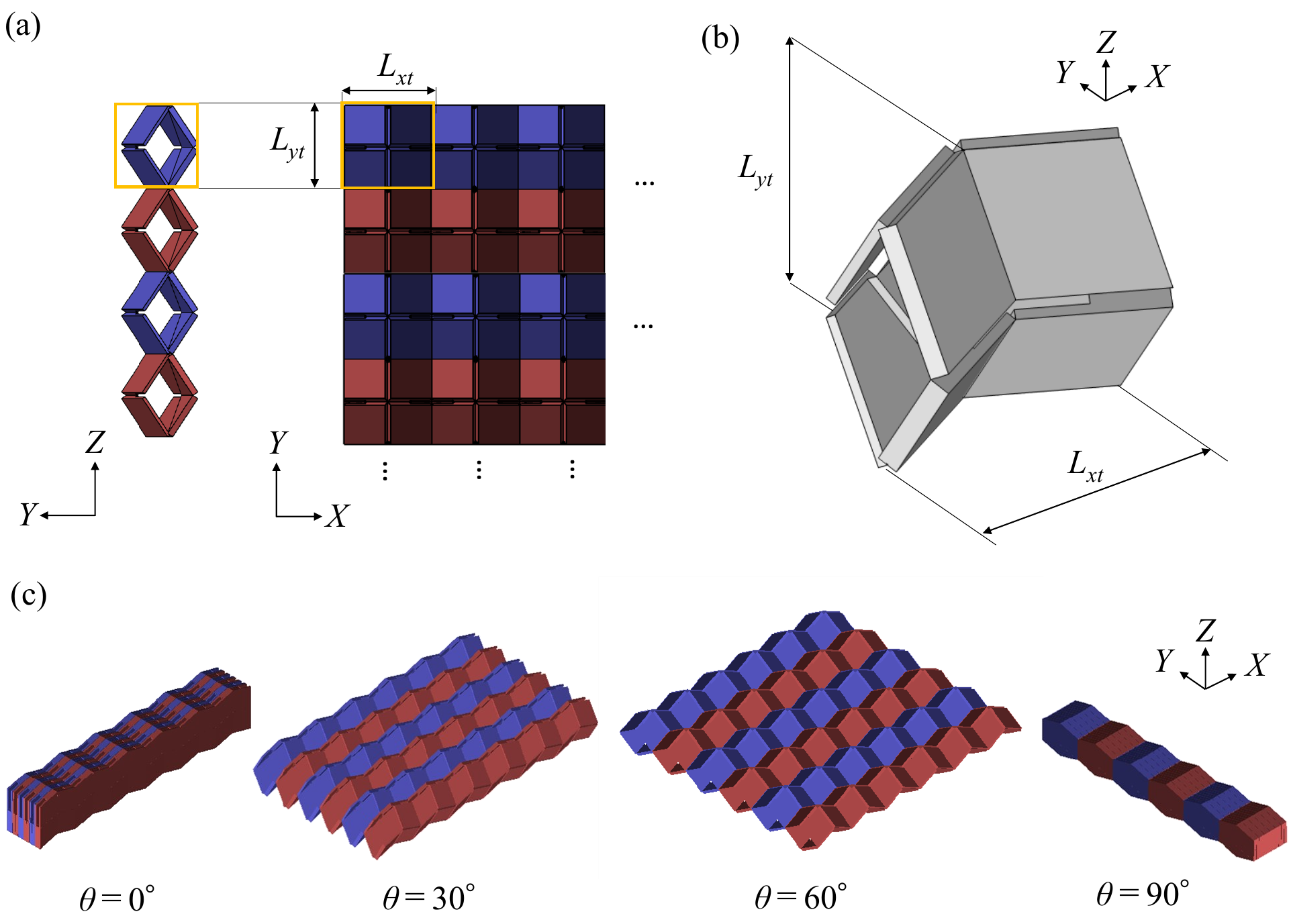}
		\caption{Tessellation of translated thick Miura-ori tubes in 2D space. (a) Coupling of the translated thick Miura-ori tubes can be tessellated by translation vector ($L_{xt}$,$L_{yt}$) on $XY$ plane. (b) Geometry of a unit cell. (c) Deformation of tessellated structures by increasing the half dihedral angle $\theta$.}
		
		\label{fig:Tessellation2D_Translation}
	\end{figure}

	By repeating coupling along the major axis (Fig.~\ref{fig:Thick Crease coupling}(a)), a tessellation of thick Miura-ori tubes in $XY$ plane was generated (Fig.\ref{fig:Tessellation2D_Translation}(a)).
	The geometry of the unit cell is illustrated in Fig.~\ref{fig:Tessellation2D_Translation}(b). 
	The size of the unit cell considering the thickness $t$ is defined as
	\begin{equation}	L_{xt}=L_x+2t\sin\beta\sin\theta,\ {\rm{where}},\ \beta 
		= \frac{V}{{L_x}/2},
		\label{eq:Lxt}
	\end{equation}
	\begin{equation}
		L_{yt}=2(H+2t\cos\theta).
		\label{eq:Lyt}
	\end{equation}
	The tessellated structures can be deformed against the half dihedral angle $\theta$ from a flat state on the $XZ$ plane to another flat state on the $YZ$ plane, as illustrated in Fig.\ref{fig:Tessellation2D_Translation}(c). 
	
	Although the tessellation of Miura-ori tubes without thickness can be stacked along the major creases using the translation vector $(0,0,2H)$ as shown in Fig.~\ref{fig:Miura tessellation}(c), stacked thick Miura-ori tubes along the seams cannot maintain one-DOF motion because the coupling will not satisfy the local mirror symmetry along the coupling interface.
	Therefore, tessellation based on coupling along the major creases is limited to two dimensions for thick Miura-ori tubes.
	
	\subsubsection{Tessellation of coupled rotated thick Miura-ori tubes}
	\label{sec:Tessellation 3D}
	
\begin{figure}[H]
		\centering\includegraphics[width=1\linewidth]{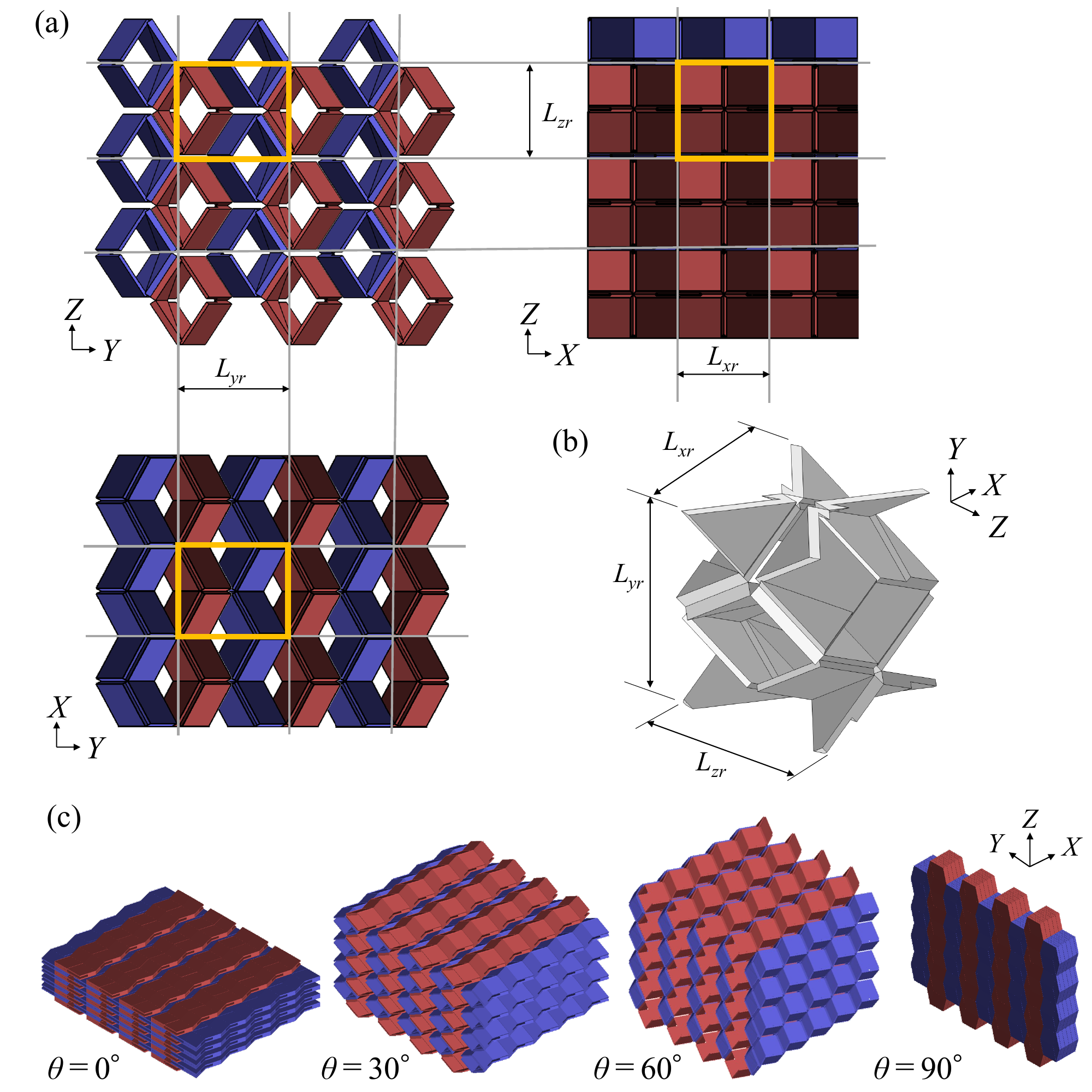}
		\caption{Tessellation by coupling the rotated thick Miura-ori tubes in 3D space. (a) Coupling of the rotated thick Miura-ori tubes can be tessellated by translation vector ($L_{xr}$,$L_{yr}$,$L_{zr}$). (b) Geometry of the unit cell. (c) Deformation of tessellated structures by increasing the half dihedral angle $\theta$.}
		
		\label{fig:Tessellation3D_Rotation}
	\end{figure}

	The coupling of the $180^\circ$-rotated Miura-ori tubes along the minor creases, as shown in Fig.~\ref{fig:Thick Crease coupling}(b) leads to tessellation in the 3D space, as shown in Fig.~\ref{fig:Tessellation3D_Rotation}(a). 
	The unit cell for tessellation is shown in Fig.~\ref{fig:Tessellation3D_Rotation}(b). The size of the unit cell considering the thickness is defined as
	\begin{equation}
		L_{xr}=L_x+2t\sin\beta\sin\theta,\ {\rm{where}},\ \beta 
		= \frac{V}{{L_x}/2} .
		\label{eq:Lxr}
	\end{equation}
	
	\begin{equation}
		L_{yr}=L_y + 2V + 2t\cos\theta\qty(\frac{2}{\cos\gamma\sin\gamma} -\tan\gamma  -\frac{2}{\tan\gamma})
		,\ {\rm{where}},\ \gamma  = \frac{H}{{L_x}/2} 
		\label{eq:Lyr}
	\end{equation}
	and
	\begin{equation}
		L_{zr}=2(H+2t\cos\theta)
		\label{eq:Lzr}
	\end{equation}
	
	The resulting cellular structures could rigidly deform from one flat state to another, as shown in Fig.~\ref{fig:Tessellation3D_Rotation}(c), because all interfaces for coupling thick Miura-ori tubes maintain local mirror symmetry. 
	This tessellation of thick Miura-ori tubes results in thickness accommodation of the interleaved cellular origami structures (Fig.~\ref{fig:Miura tessellation}(d)) proposed in \cite{RN11}. 
	Within the scope of this study, this was the only 3D tessellation of Miura-ori tubes with thickness.
	
	\subsubsection{Tessellation of coupled mirrored thick Miura-ori tubes}
	\label{sec:Tessellation 2D Mirror}
\begin{figure}[H]
		\centering\includegraphics[width=1\linewidth]{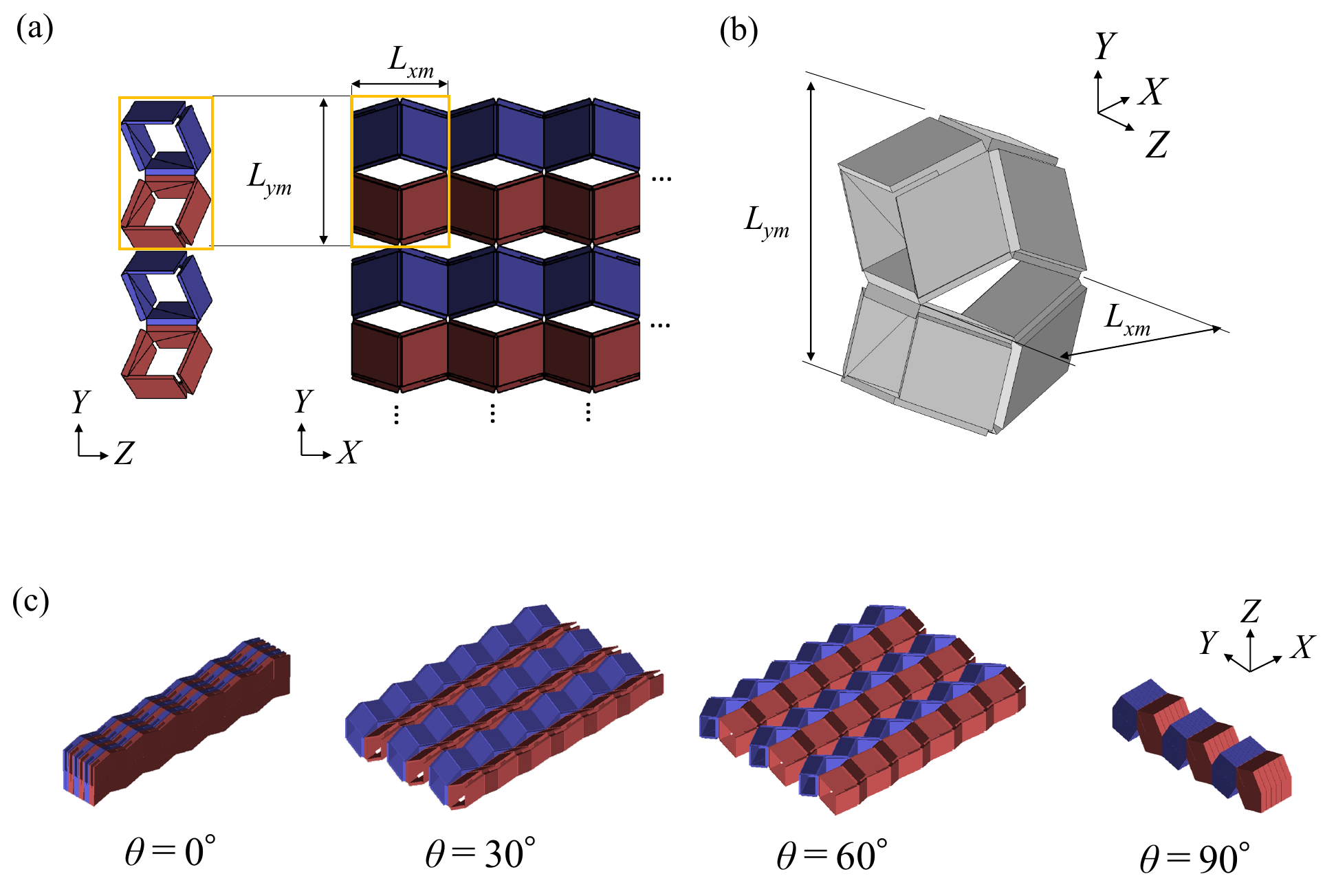}
		\caption{Tessellation of the coupling of thick mirrored Miura-ori tubes in 2D space. (a) Coupling of the thick Miura-ori tubes can be tessellated by translation vector ($L_{xm}$,$L_{ym}$) on the $XY$ plane. (b) Geometry of unit cell. (c) Deformation of tessellated structures by increasing the half dihedral angle $\theta$.}
		
		\label{fig:Tessellation2D_Mirror}
	\end{figure}

	The thick mirrored Miura-ori tubes, shown in Fig.~\ref{fig:Thick Crease coupling}(c), can be tessellated along the minor creases on the $XY$ plane, as indicated in Fig.~\ref{fig:Tessellation2D_Mirror}(a), by translation, with lengths $L_{xm}$ and $L_{ym}$ defined as
	\begin{equation}
		L_{xm}=L_x+2t\sin\beta\sin\theta,\ {\rm{where}},\ \beta 
		= \frac{V}{{L_x}/2} .
		\label{eq:Lxm}
	\end{equation}
	\begin{equation}
		L_{ym}=2 \sin\gamma\qty[L_y + V + t \cos\theta\qty(\frac{2}{\cos\gamma\sin\gamma}-\tan\gamma ) ]
		,\ {\rm{where}},\ \gamma  = \frac{H}{{L_x}/2} 
		\label{eq:Lym}
	\end{equation}
	The geometry of the unit cell is illustrated in Fig.~\ref{fig:Tessellation2D_Mirror}(b). The tessellated structures can be deformed, as shown in Fig.~\ref{fig:Tessellation2D_Mirror}(c).
	The stacking of this tessellation cannot maintain a one-DOF motion because it contains a face contact, as shown in Fig.~\ref{fig:Miura tessellation}(e).
	Therefore, tessellation based on mirror coupling by a minor-crease contact is limited to two dimensions for thick Miura-ori tubes.
	
	\subsection{Variation of geometries}
	\subsubsection{Combination of couplings of translated and mirrored tubes}
\begin{figure}[H]
		\centering\includegraphics[width=1\linewidth]{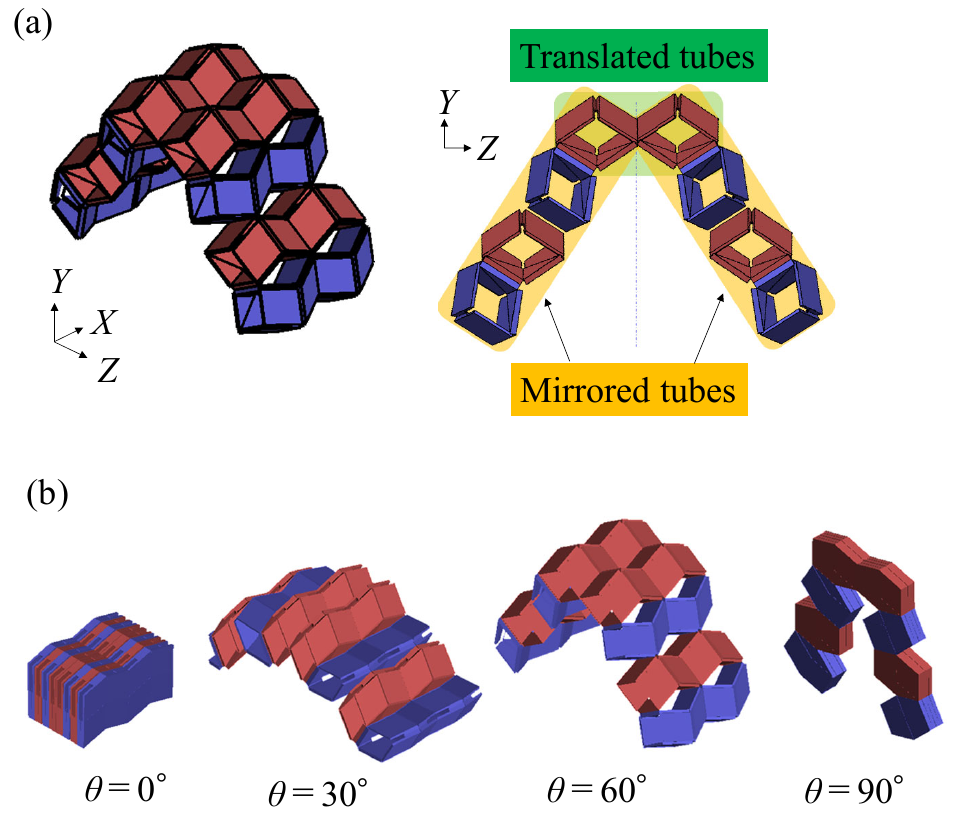}
		\caption{Combination of translated and mirrored Miura-ori tubes. (a) The geometry of arch shapes formed by the coupling of thick Miura-ori tubes with $\theta=60^\circ$. (b) Arch is deployed by increasing the half dihedral angle $\theta$.}
		\label{fig: Arch}
	\end{figure}

	The combination of different types of coupled Miura-ori tubes results in various motions and geometries. For instance, the combination of coupling by the minor- and major-crease contacts forms arch shapes, as shown in Fig.\ref{fig: Arch}(a). This mechanism can deploy arch shapes from flat folded shapes, as shown in Fig.\ref{fig: Arch}(b) as the half dihedral angle $\theta$ increases. 
	
	\subsubsection{Coupling of graded Miura-ori tubes}
\begin{figure}[H]
		\centering\includegraphics[width=1\linewidth]{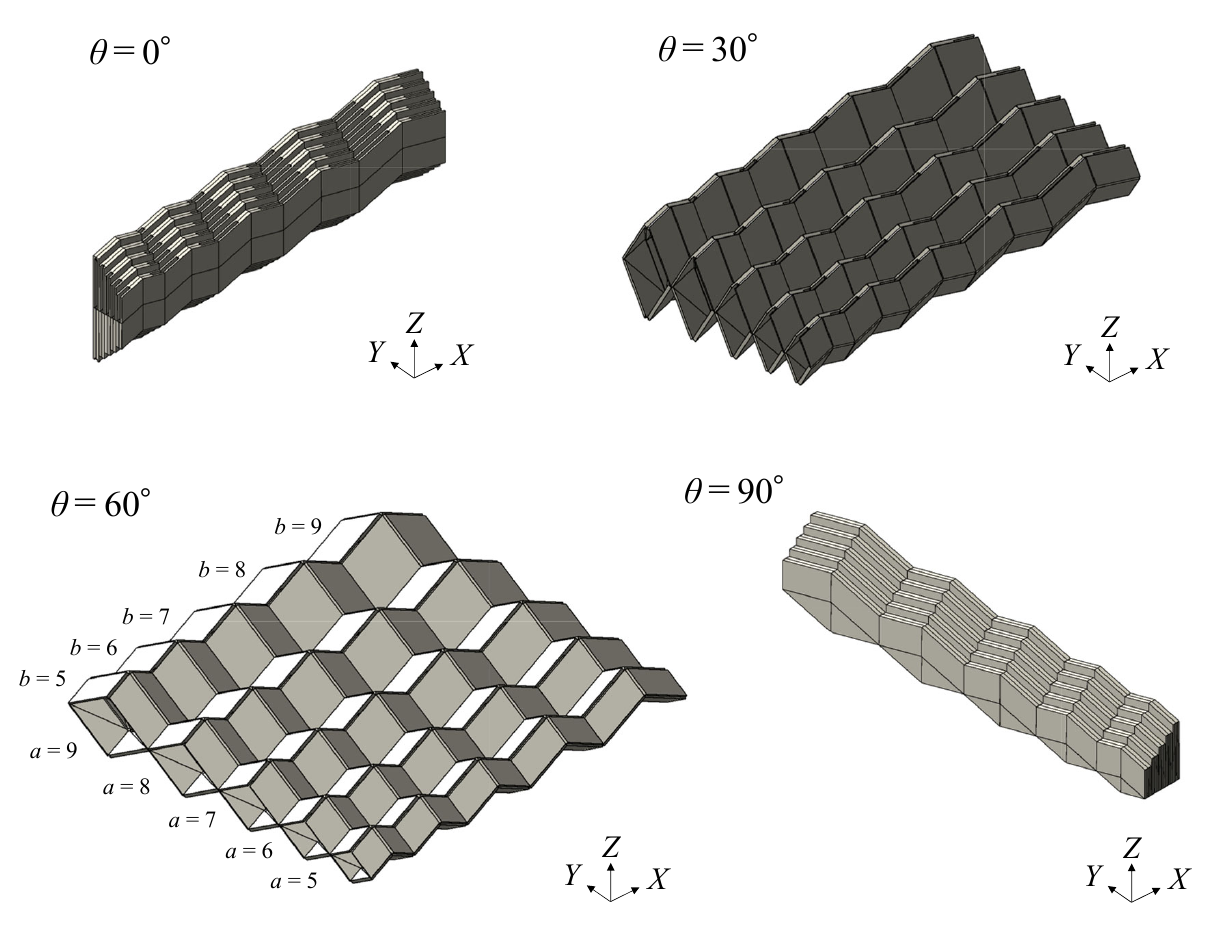}
		\caption{Coupling of thick Miura-ori tubes with lengths $a$ and $b$ graded from 5 to 9 along the $X$- and $Y$-axes. The coupled thick Miura-ori tubes are flat-foldable}
		\label{fig: Graded}
	\end{figure}

	Thick Miura-ori tubes can be coupled along the creases if local mirror symmetries around the coupling interfaces are maintained.
	Therefore, the dimensions of the tube parallelograms need not be identical.
	As an example of coupling different Miura-ori tubes, graded Miura-ori tubes were coupled as shown in Fig.~\ref{fig: Graded}, where the lengths of parallelograms $a$ and $b$ are graded from five to nine along the $x$- and $y$-axes, respectively. 
	The graded Miura-ori tubes can be deformed from a flat state on the $XZ$ plane to another flat state on the $YZ$ plane.
	
	\section{Stiffness of coupled Miura-ori tubes}
	\label{sec:Elastic deformation}
	
	\subsection{Bar-and-hinge models for eigenvalue analysis}
	\label{sec:Bar-and-hinge}
	To realize stiff deployable structures, soft deformation caused by one-DOF mechanism and high stiffness for elastic deformation are required. This ideal mechanical property of stiff deployable structures is characterized by wide eigenvalue gaps between the low eigenvalues associated with one-DOF deformations and the high eigenvalues associated with elastic deformations. Therefore, eigenvalue gaps can be used as indicators of stiff deployable structures \cite{RN13}. Because the eigenvalue analysis under the free boundary condition provides trivial modes from the first to the sixth eigenmodes, the eigenmode caused by the one-DOF deformation occurs as the seventh mode, and the eigenmode caused by elastic deformation occurs higher than the seventh mode.
	
\begin{figure}[H]
		\centering\includegraphics[width=0.8\linewidth]{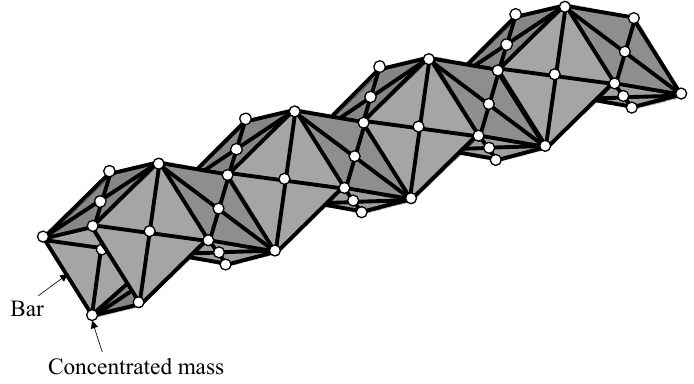}
		\caption{Schematic of bar-and-hinge models. The panels are discretized using five nodes and eight bars.}
		\label{fig:bar and hinge}
	\end{figure}

	We performed an eigenvalue analysis of the bar-and-hinge model to evaluate the stiffness of the coupled Miura-ori tubes.
	The eigenvalue problem was defined using the stiffness and mass matrices $\bf{K}$ and $\bf{M}$ as follows:
	\begin{equation}
		{\bf{K}}{\bf{v}}_i=\lambda_i{\bf{M}}{\bf{v}}_i.
		\label{eq:Eigenvalue problem}
	\end{equation}
	The stiffness and mass matrices were constructed using bar-and-hinge models formulated using the N5B8 approach, which expresses their panels using five nodes and eight bars, as shown in Fig.~\ref{fig:bar and hinge}. 
	Eigenvalues $\lambda_i$ and eigenmodes ${\bf{v}}_i$ are associated with the stiffness of each deformation.
	The mass matrix $\bf{M}$ is defined as a diagonal matrix composed of concentrated masses, assuming that the masses of the triangle panels are equally distributed on each node. 
	The stiffness matrices of the bar-and-hinge models were implemented based on MERLIN, which was developed by \cite{RN195,RN197}. 
	The bar-and-hinge model does not geometrically reflect the thickness accommodation suggested in the previous section; instead, the structural effect of thickness can be modeled as the bar-and-hinge stiffness.
	
	The geometry of the Miura-ori tube is defined by the parameters $a=1$, $c=1$ and $\alpha=55^\circ$ \cite{RN13}. The Young’s modulus, density, Poisson’s ratio, and thickness of the panels were $10^6$, 1, 1/3, and 0.01, respectively, which were borrowed from \cite{RN13}. The scaling factor for the stiffness of the hinges along the creases was set to 40, as in \cite{RN17}. As the extension was defined by $L_{x}/2c\times100\ [\%]$, the eigenvalues were evaluated for each expansion state. An eigenvalue analysis of the dynamic systems defined by the bar-and-hinge models was implemented using MATLAB R2022b.

	\subsection{Formation of eigenvalue gaps by coupling of Miura-ori tubes}
	\label{sec:Bar-and-hinge result}
\begin{figure}[H]
		\centering\includegraphics[width=1\linewidth]{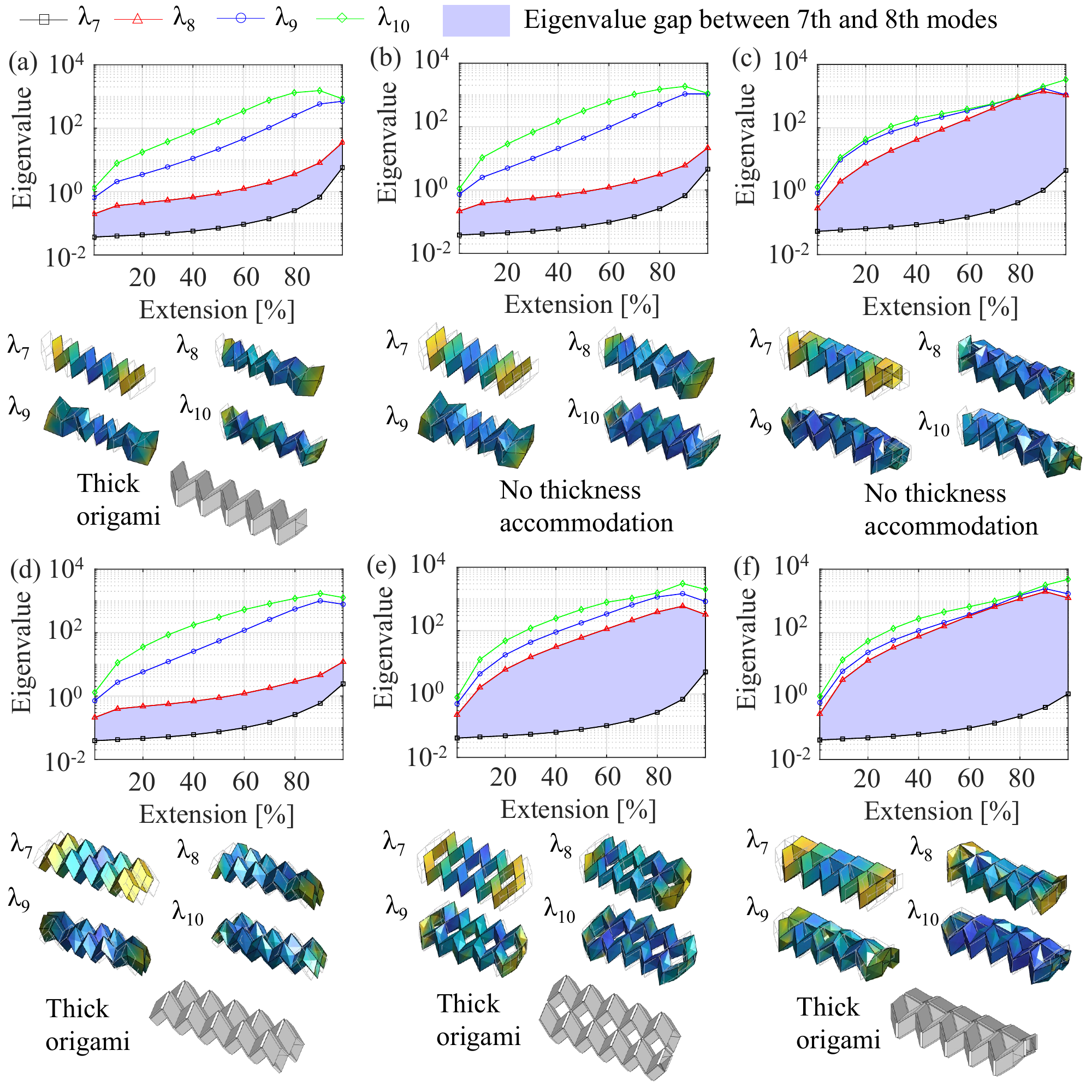}
		\caption{Eigenvalues of the coupled Miura-ori tubes defined by bar-and-hinge models. (a) Single Miura-ori tube (b) Face contact by translation. (c) Face contact by glide reflection (i.e., zipper-coupled tubes). (d) Major-crease contact by translation. (e) Minor-crease contact by rotation. (f) Minor-crease contact by mirror.}
		\label{fig:Eigenvalue}
	\end{figure}

	Miura-ori tubes have eigenmodes with soft deformation characterized by squeezing of the cross-section of the tubes, in addition to the one-DOF modes \cite{RN13}. 
	Thus, the eigenvalue gaps between the seventh and eighth modes were small, as indicated in Fig.~\ref{fig:Eigenvalue}(a). 
	The squeezing modes must be eliminated to achieve high stiffness throughout deployment.
	As discussed in previous studies \cite{RN194}, although the coupling of translated Miura-ori tubes along their faces cannot widen the eigenvalue gaps between the seventh and eighth modes, as shown in Fig.~\ref{fig:Eigenvalue}(b), the zipper coupling can provide wide eigenvalue gaps, as shown in Fig.~\ref{fig:Eigenvalue}(c), owing to the elimination of squeezing modes.
	This result indicates that the zipper coupling increases the stiffness of elastic deformation by two orders of magnitude compared to the single tube or Miura-ori tubes coupled along the faces by translation, reproducing the results of \cite{RN13}.
	Although zipper-coupled tubes offer high stiffness throughout the deployment, coupling by face contact does not provide thickness accommodation because of the constraints caused by the thickness of the panels.
	
	To solve the thickness issues, the coupling of translated (major-crease contact), rotated (minor-crease contact), and mirrored (minor-crease contact) Miura-ori tubes was investigated, as shown in Fig.~\ref{fig:Eigenvalue}(d)–(f), respectively. 
	As observed in Fig.~\ref{fig:Eigenvalue}(d), although the coupling of the translated Miura-ori by the major-crease contact cannot widen the eigenvalue gap, the coupling of the rotated and mirrored Miura-ori tubes by the minor-crease contact leads to wide eigenvalue gaps (Fig.~\ref{fig:Eigenvalue}(e) and (f)), respectively.
	These eigenvalue gaps had approximately the same ranges as those obtained by zipper coupling.
	Therefore, coupling the rotated or mirrored Miura-ori tubes by minor-crease contact enables thickness accommodation without loss of the unique mechanical properties of the zipper-coupled tubes, which increase the stiffness by two orders of magnitude.
	
	\subsection{Eigenvalue gaps of tessellated structures}
	\subsubsection{Stiffness of plate-like structures by 2D tessellation}
\begin{figure}[H]
		\centering\includegraphics[width=1\linewidth]{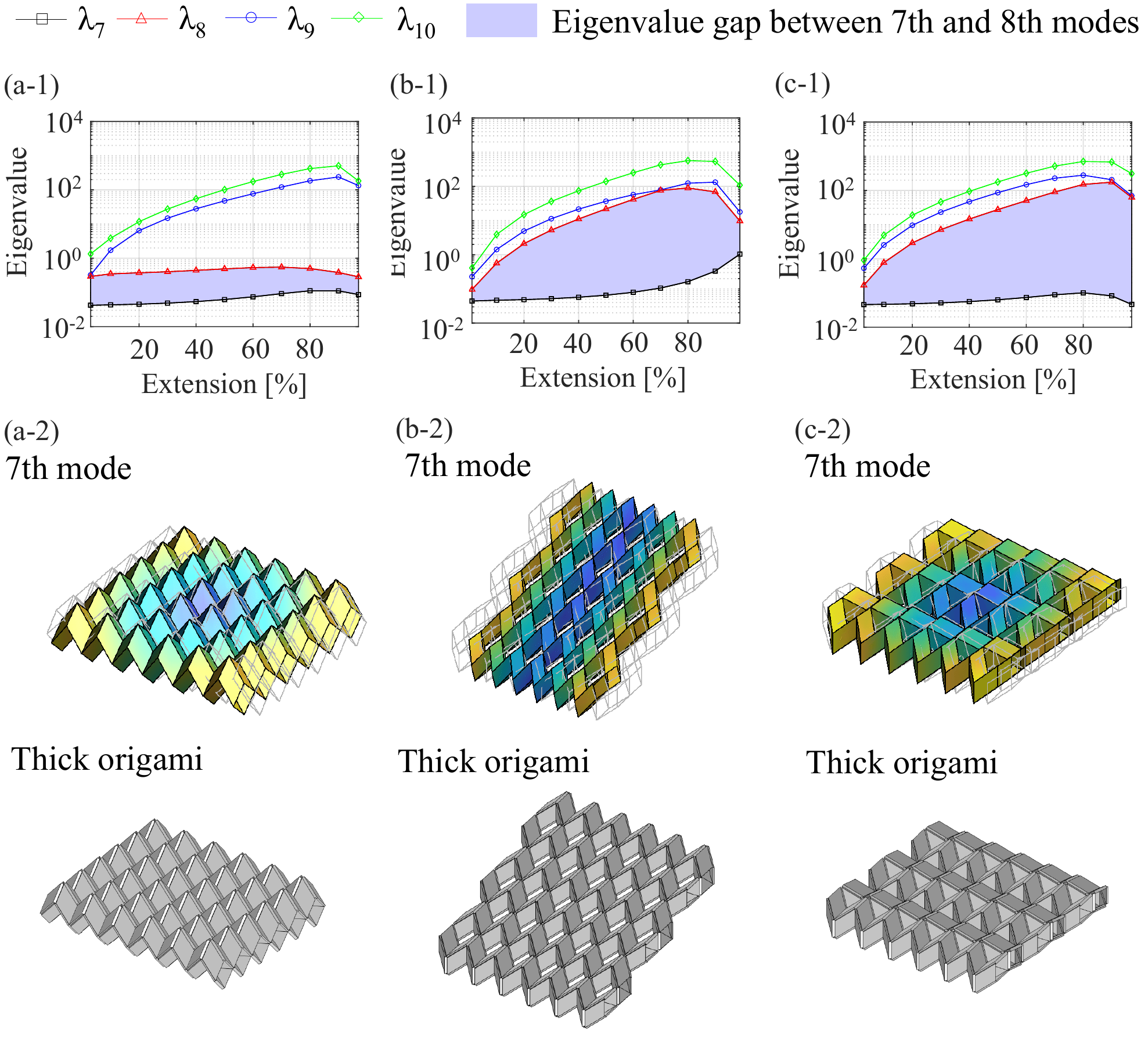}
		\caption{Eigenvalues of tessellated structures in 2D space. (a-1) Major-crease contact by translation. (b-1) Minor-crease contact by mirror. (c-1) Minor-crease contact by rotation. (b-1)-(b-3) Seventh mode and thickness accommodation to retain the seventh mode after tessellation.}
		\label{fig:Eigenvalue of 2D}
	\end{figure}

	To investigate the stiffness of the tessellated structures in 2D space, an eigenvalue analysis of the structures using translated, rotated, and mirrored Miura-ori tubes that can be coupled by major- and minor-crease contact was performed, as shown in Fig.~\ref{fig:Eigenvalue of 2D}.
	As illustrated in Fig.~\ref{fig:Eigenvalue of 2D}(a-1), the coupling of the translated Miura-ori tubes by major-crease contact maintained a small eigenvalue gap, which was observed as the virtual DOF of the tessellated Miura-ori tubes \cite{RN10}.
	In contrast, the tessellation of rotated or mirrored Miura-ori tubes by a minor-crease contact provided wide eigenvalue gaps, as shown in Fig.~\ref{fig:Eigenvalue of 2D}(b-1) and (c-1), respectively. Therefore, these tessellations by minor-crease contact can increase the stiffness of the plate-like structures compared with the coupling of the translated Miura-ori tubes by major-crease contact.
	In addition, all tessellated structures maintained one-DOF motion when thick Miura-ori tubes were coupled along the creases, as shown in Fig.~\ref{fig:Eigenvalue of 2D}(a-2)–(c-2). 
	Therefore, the coupling of rotated or mirrored tubes by a minor-crease contact has the advantage of achieving stiff, deployable plate-like structures using thick panels.
	
	\subsubsection{Stiffness of cellular structures by 3D tessellation}
\begin{figure}[H]
		\centering\includegraphics[width=0.9\linewidth]{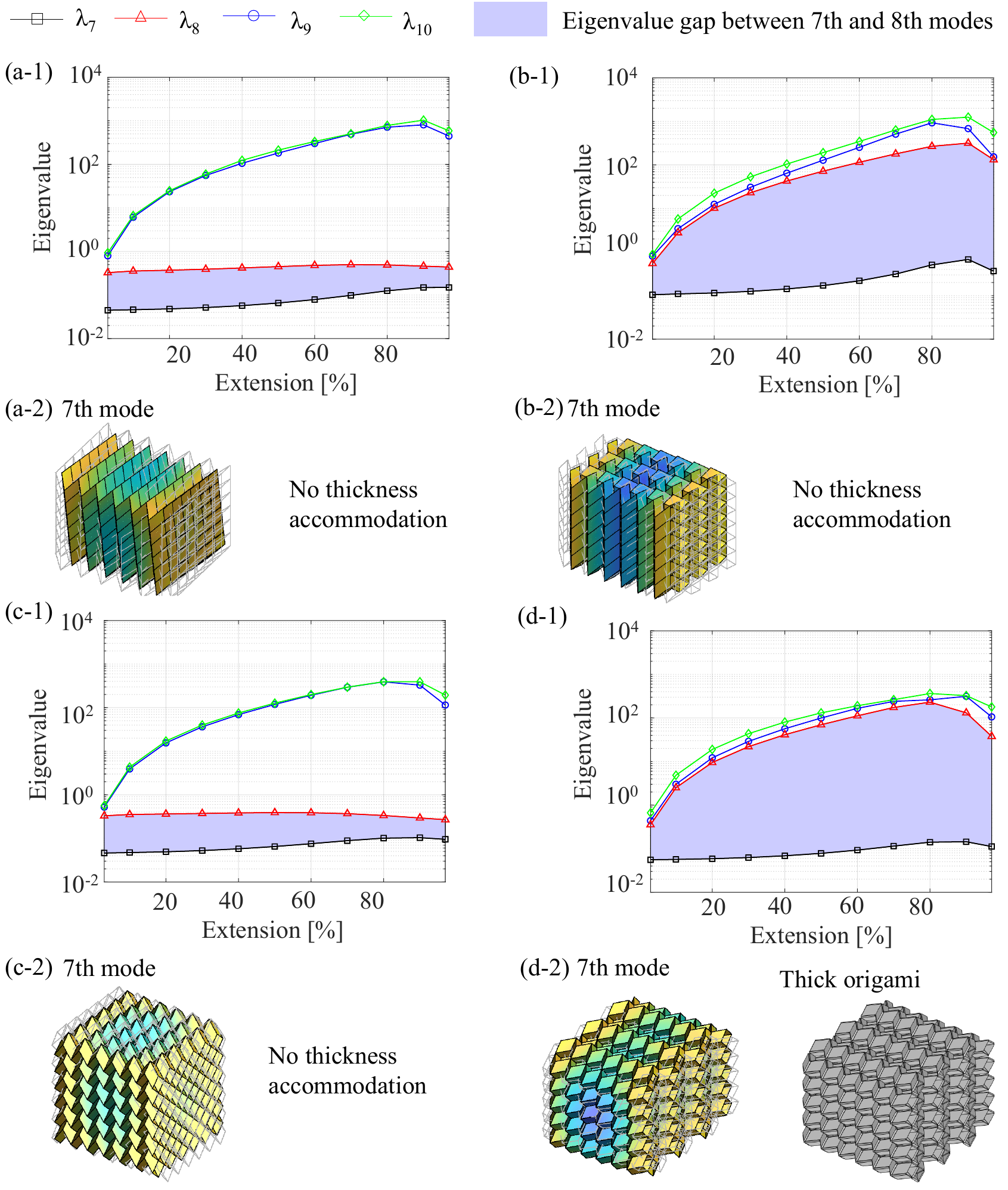}
		\caption{Eigenvalues of cellular structures by tessellation of Miura-ori tubes. (a-1) Face-contact translation (b-1) Face-contact glide-reflection (i.e., Zipper coupling) (c-1) Combination of the seam and major-crease contact. The tessellation is equivalent to the tessellated structure shown in (a-1) (only the boundary is different). (d-1) Combination of major- and minor-crease contact, which can keep one-DOF motion via coupling of thick Miura-ori tubes. (a-2)-(d-2) Seventh mode of the cellular structures.
			Only coupling of rotated Miura-ori tubes by minor-crease contact can retain the seventh mode after thickness accommodation.}
		\label{fig:Eigenvalue_Cell}
	\end{figure}

	Cellular structures formed by the tessellation of Miura-ori tubes were evaluated using bar-and-hinge models.
	The eigenvalues of the tessellated structures obtained by coupling the translated and rotated Miura-ori tubes via face contact are shown in Fig.~\ref{fig:Eigenvalue_Cell}(a-1) and (b-1). 
	The eigenvalues of the tessellated structure obtained by coupling the translated and rotated Miura-ori tubes with edge contacts are shown in Fig.~\ref{fig:Eigenvalue_Cell}(c-1) and (d-1). 
	The tesselations used for the eigenvalue analysis shown in Fig.~\ref{fig:Eigenvalue_Cell}(a-1) and (c-1) are equivalent; except for the boundaries of the tessellated structures, the eigenvalue gaps are approximately equal to and smaller than one order of magnitude. 
	The rotated Miura-ori tubes along the faces (i.e., zipper coupling) provide wider eigenvalue gaps compared to the coupling based on translation, where the eigenvalue gaps are approximately two orders of magnitude.
	The coupling of the rotated tubes by the minor-crease contact yielded similarly wide eigenvalue gaps.
	
	All cellular structures exhibited a rigid folding motion, as indicated by the seventh mode in Fig.~\ref{fig:Eigenvalue_Cell}(a-2)–(d-2), if the panel thickness is not considered. 
	However, coupling only the rotated thick Miura-ori tubes by minor-crease contact can provide thickness accommodation for 3D tessellation, as explained in Section~\ref{sec:Geometry}. 
	From these results, tessellation by coupling the rotated thick Miura-ori tubes by minor-crease contact facilitates not only a valid mechanism using thick panels but also high stiffness of cellular structures throughout the deployment.
	
	\subsection{Thick-panel structures with stiffness}
	In summary, the coupling structures of thick Miura-ori tubes exhibit high stiffness (two orders of magnitude) while preserving their mechanisms when coupled with mirrored or rotated tubes along minor creases (Fig.~\ref{fig:Eigenvalue} (e) and (f)).
	Both coupling reactions allowed planar tessellation (Fig.~\ref{fig:Eigenvalue of 2D} (b-1) and (c-1)), whereas only the coupling of rotated tubes along the minor creases allows for spatial tessellation (Fig.~\ref{fig:Eigenvalue_Cell} (d-2)).

	\section{Meter-scale construction}
	\label{sec:fabrication}
\begin{figure}[H]
		\centering\includegraphics[width=0.9\linewidth]{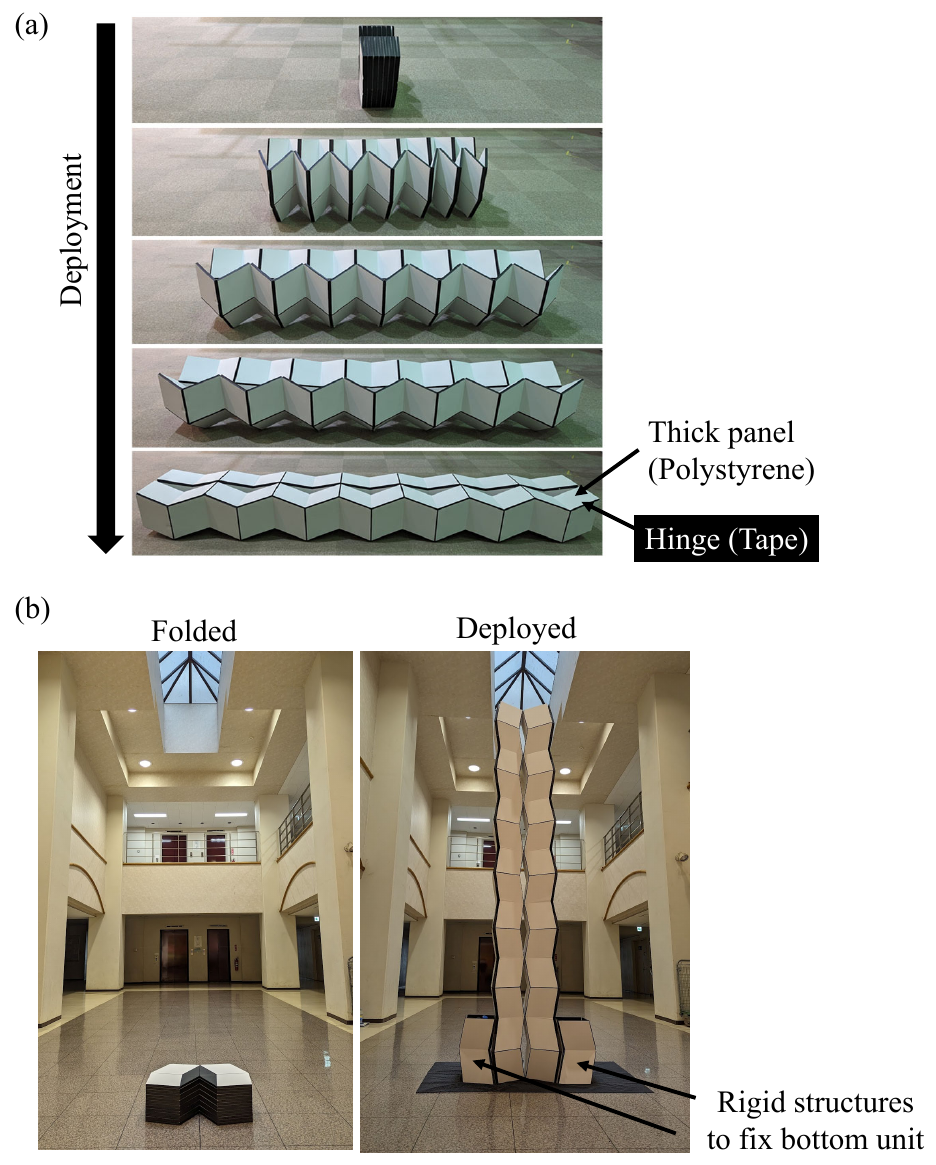}
		\caption{Meter-scale fabrication of the coupled thick mirrored Miura-ori tubes. (a) Deployment motion of the coupled thick Miura-ori tubes. The coupled tubes can be folded via one-DOF motion. (b) Construction of meter-scale pillar by deployment of the coupled thick Miura-ori tubes. The coupled Miura-ori tubes are stabilized by the constraints of the bottom units.}
		\label{fig:CoupledPillar}
	\end{figure}

	To demonstrate the coupled thick Miura-ori tubes along their creases, coupled mirrored Miura-ori tubes along the minor creases, which have high stiffness, were fabricated on a meter scale. 
	The unit cell of the Miura-ori tubes was defined by the lengths of the parallelograms $a=230\ \rm{mm}$, $b=230\ \rm{mm}$ and the internal angle of the parallelogram $\alpha=75^\circ$. 
	The panels were fabricated by cutting a 7-mm-thick expanded polystyrene board (JSP Corporation), assembled using rendering tape (3M Company), and covered with films (3M Company). 
	The tapes and films behaved as hinges on the folding lines. 
	The fabricated structures were flat-foldable and deployable via one-DOF motion, as shown in Fig.~\ref{fig:CoupledPillar}(a). 
	
	A stiff pillar was constructed using coupled thick Miura-ori tubes, as shown in Fig.~\ref{fig:CoupledPillar}(b). 
	The construction process is as follows: (1) deploy the coupled tubes in the horizontal direction, (2) rotate and stand them in the vertical orientation, and fix the bottom units by placing two rigid support structures in which the water tank is filled.  
	The constructed pillar is stable because fixing the bottom unit can adequately stabilize the entire structure owing to the high stiffness demonstrated in the previous section.
	We direct the reader to the supplementary video files for a more detailed description of the motion of the pillar constructed by deployment.
	The fabricated structure suggests that the proposed coupling contributes to high transportability by flat foldability and rapid construction of stiff structures by one-DOF motions.
	
	\section{Conclusion}
	\label{Sec:Conclusion}
	This study proposed coupled thick Miura-ori tubes along the creases to maintain one-DOF deformation by local mirror symmetry around the coupling interfaces. The findings of this study are as follows: 
	\begin{itemize}
		\item The coupling of mirrored or rotated Miura-ori tubes along minor creases and the coupling of translated Miura-ori tubes along major creases were identified as the coupling of Miura-ori tubes that can accommodate thickness.
		\item The coupling of rotated Miura-ori tubes along minor creases leads to thickness accommodation of spatial tessellation structures consisting of thick Miura-ori tubes. This can be interpreted as the thickness accommodation of interleaved origami cellular structures.
		\item The coupling of mirrored or rotated Miura-ori tubes along minor creases increases the stiffness of the coupled Miura-ori tubes by two orders of magnitude after coupling, which is the same result as that for the zipper coupled tubes.
		\item A meter-scale deployable pillar was constructed using thick mirrored Miura-ori tubes by one-DOF motions and the high stiffness of the coupled Miura-ori tubes.
		
	\end{itemize}
	These findings suggest that the proposed structures have potential applications for the rapid construction of structures by one-DOF motion and enhancement of transportability via flat foldability.
	
	Although this study focused on Miura-ori tubes, the proposed approach could be applied to the coupling of different thick origami-based structures with local mirror symmetries along the coupling interfaces, which could enable various tessellated structures and mechanical properties.
	In addition, fabrication processes such as additive manufacturing \cite{RN456} can be explored to construct tessellated structures more efficiently by coupling thick origami-based structures.
	
	\newpage
	
	\appendix
	\setcounter{figure}{0}
	\section{\label{sec:Motion}Motion of coupled Miura-ori tubes}
\begin{figure}[H]
		\centering\includegraphics[width=0.9\linewidth]{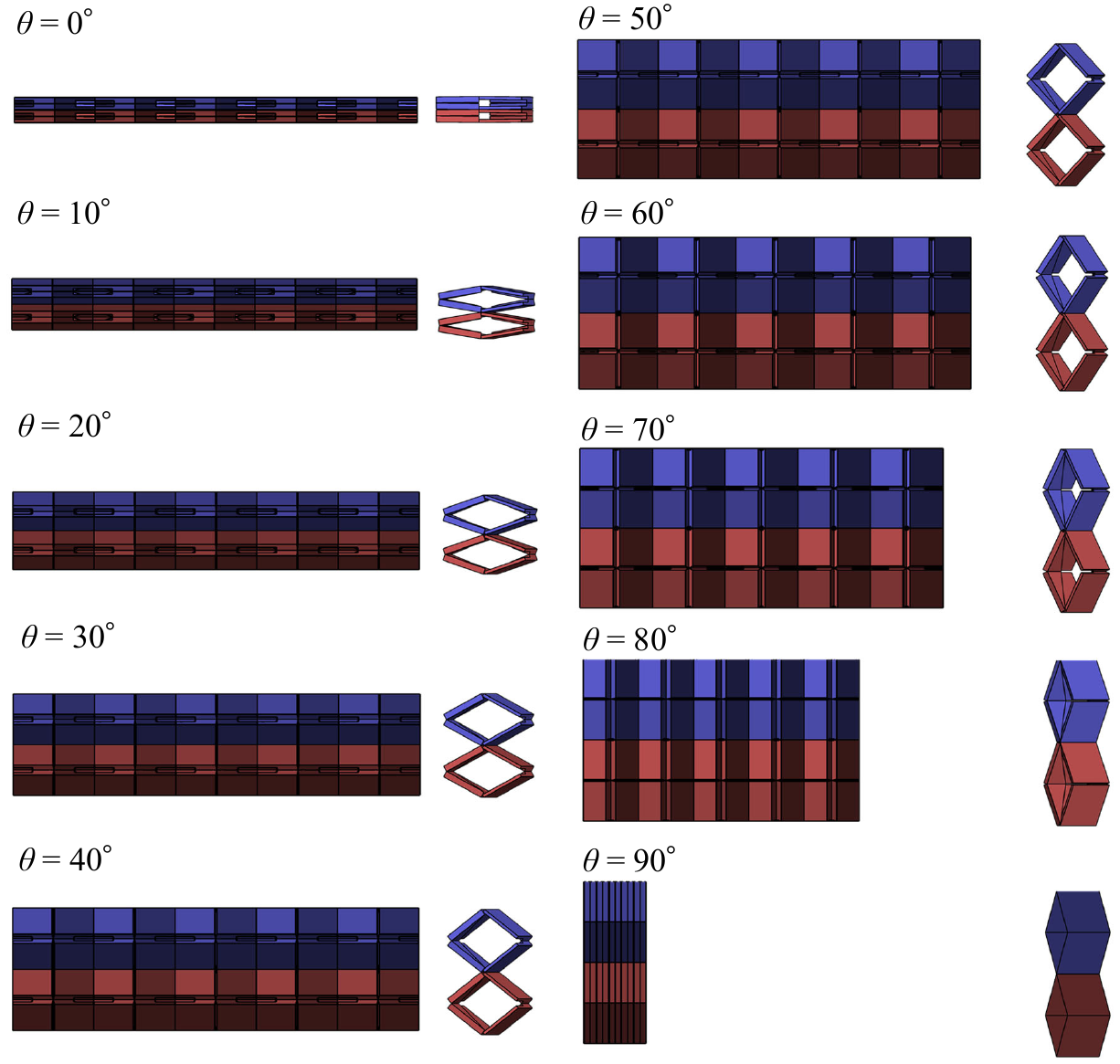}
		\caption{Folding motion of the coupled translated thick Miura-ori tubes}
		\label{fig:Motion of Translated tubes}
	\end{figure}

	Motion due to the coupling of translated Miura-ori tubes by major-crease contact is illustrated in Fig.~\ref{fig:Motion of Translated tubes}.
	
	\newpage

\begin{figure}[H]
		\centering\includegraphics[width=1\linewidth]{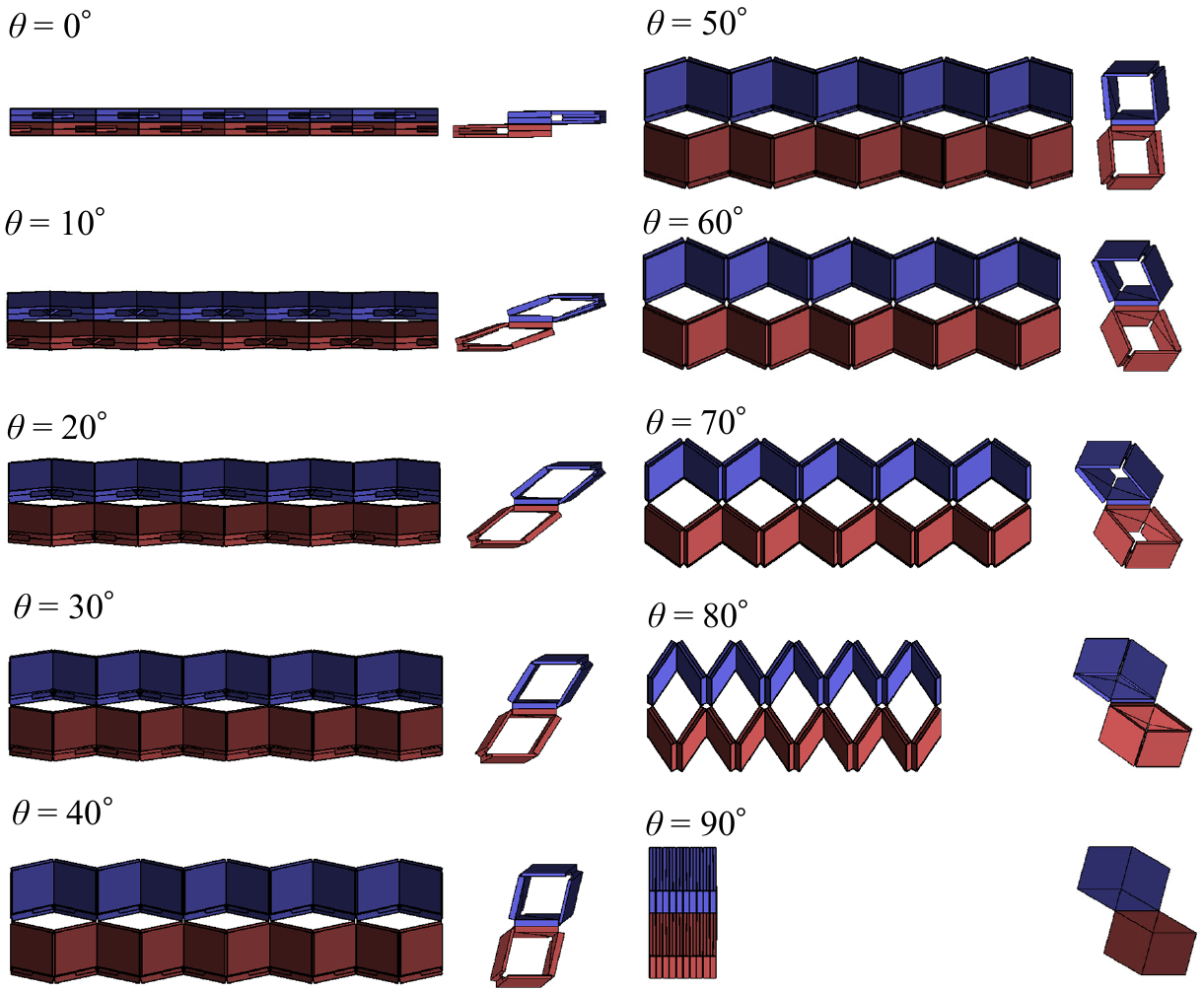}
		\caption{Folding motion of the coupling rotated thick Miura-ori tubes}
		\label{fig:Motion of rotated tubes}
	\end{figure}

	Motion due to the coupling of rotated Miura-ori tubes by minor-crease contact is depicted in Fig.~\ref{fig:Motion of rotated tubes}.
	
	\newpage
	
\begin{figure}[H]
		\centering\includegraphics[width=1\linewidth]{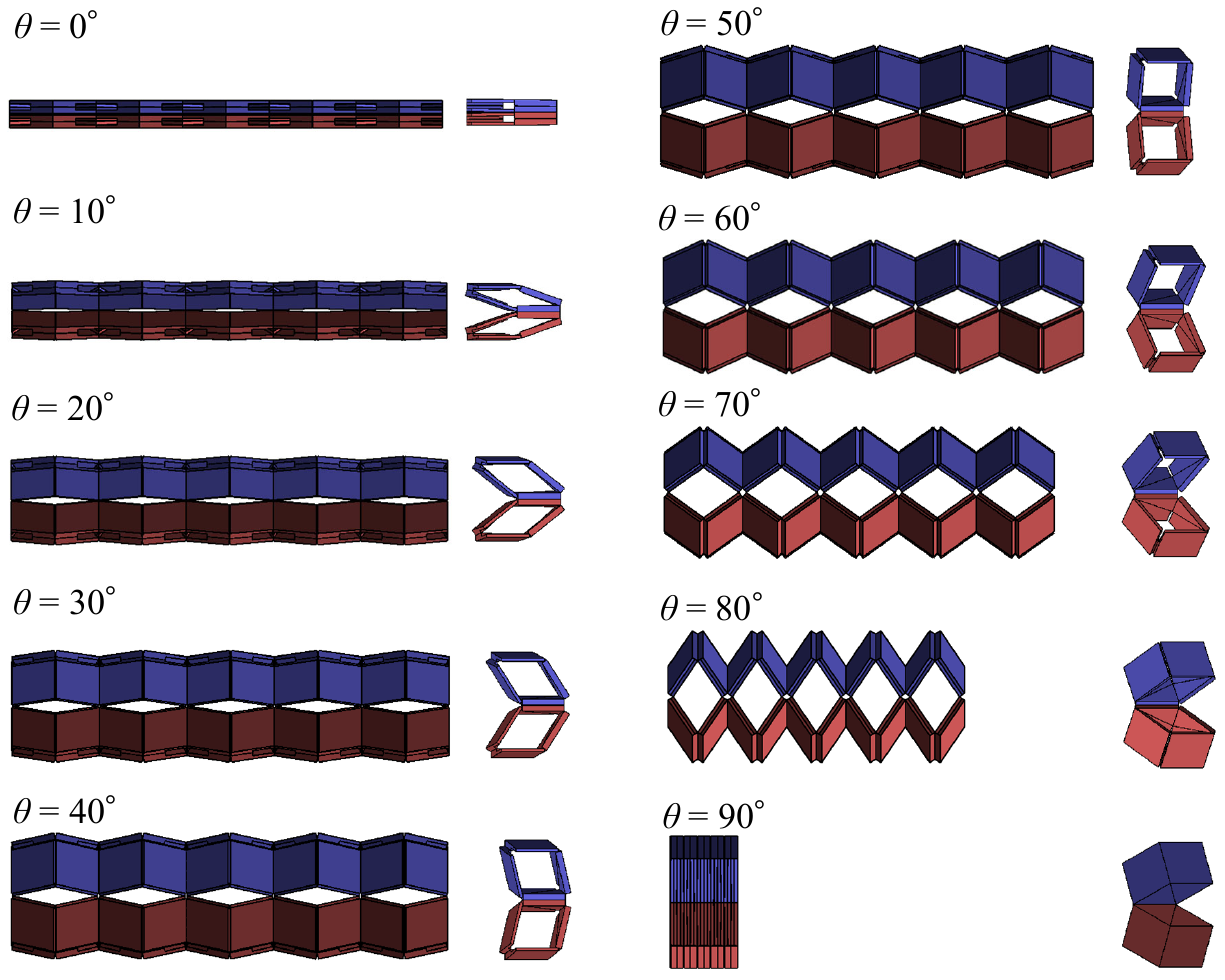}
		\caption{Folding motion of the coupled mirrored thick Miura-ori tubes}
		\label{fig:Motion of mirrored tubes}
	\end{figure}

	Motion due to the coupling of the mirrored Miura-ori tubes by minor-crease contact is shown in Fig.~\ref{fig:Motion of mirrored tubes}.
	
	\newpage
	\section*{\label{sec:Videos}Supplementary material}
	The supplementary data for this study are illustrated as follows
	
	\begin{figure}[H]
		\begin{center}
			\begin{tabular}{c}
				\includegraphics[clip, width=0.7\linewidth]{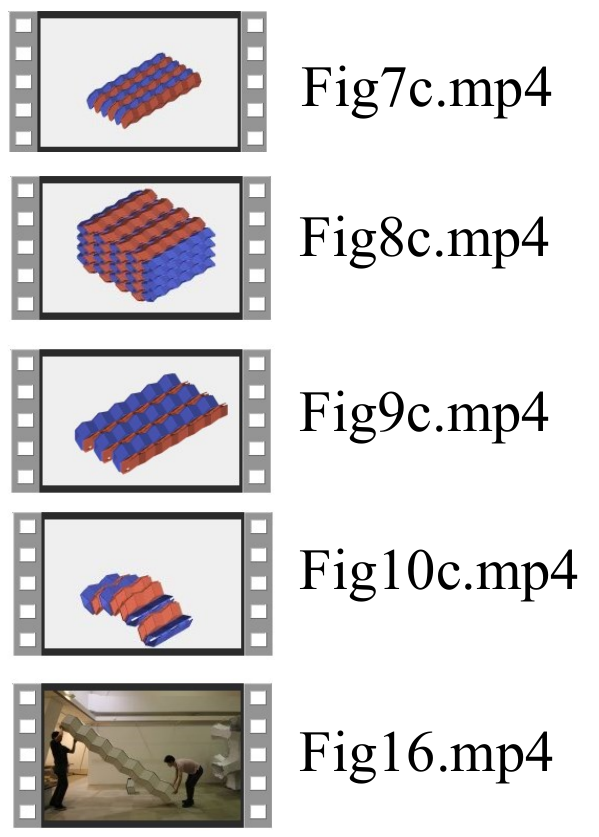}
			\end{tabular}
		\end{center}		
	\end{figure}
	
	\section*{Declaration of Competing Interest}
	The authors declare that they have no competing financial interests or personal relationships that may have influenced the work reported in this study.
	
	\section*{Acknowledgments}
	This study did not receive any specific grants from funding agencies in the public, commercial, or non-profit sectors.

	\bibliographystyle{elsarticle-num} 
	\bibliography{OrigamiBib}
	
	
		
		
		
\end{document}